\documentclass{article}
\usepackage[utf8]{inputenc}

\usepackage{amsmath}

\usepackage{cite}
\usepackage{amsmath,amssymb,amsfonts}
\usepackage{algorithmic}
\usepackage{geometry}
\usepackage{graphicx}
\usepackage{textcomp}
\usepackage{xcolor}
\usepackage{amsthm}
\usepackage{subcaption}

\def\BibTeX{{\rm B\kern-.05em{\sc i\kern-.025em b}\kern-.08em
    T\kern-.1667em\lower.7ex\hbox{E}\kern-.125emX}}
\begin{document}

\title{Comparative Sentiment Analysis of  App Reviews
 }

\author{
\begin{tabular}{cc}
   \large Sakshi Ranjan$^1$ & \large Subhankar Mishra$^{2,3}$ \\
    &\\
    \multicolumn{2}{c}{\textit{${^1}$Department of Computer Science and Applications, Utkal University}}\\
    \multicolumn{2}{c}{Bhubaneswar - 751004, India}\\
    &\\
    \multicolumn{2}{c}{\textit{${^2}$School of Computer Sciences,}}\\
    \multicolumn{2}{c}{\textit{National Institute of Science, Education and Research Bhubaneswar}}\\
    \multicolumn{2}{c}{India - 752050}\\
    &\\
    \multicolumn{2}{c}{\textit{${^3}$Homi Bhaba National Institute, Anushaktinagar, Mumbai}}\\
     \multicolumn{2}{c}{India - 400094}\\
     &\\
     \multicolumn{2}{c}{${^1}$sakshi.ranjan07@gmail.com }\\
     \multicolumn{2}{c}{$^{2}$ smishra\@niser.ac.in}\\
     \end{tabular}
}
% \author{\IEEEauthorblockN{Sakshi Ranjan}
% \IEEEauthorblockA{\textit{Department of Computer Science and Applications} \\
% \textit{Utkal University}\\
% Bhubaneswar - 751004, India\\
% sakshi.ranjan07@gmail.com \\ORCID: 0000-0002-1740-8366}
% \and
% \IEEEauthorblockN{Subhankar Mishra}
% \IEEEauthorblockA{\textit{School of Computer Sciences} \\
% \textit{National Institute of Science Education and Research}\\
% Bhubaneswar, India - 752050 \\
% \textit{Homi Bhabha National Institute}, \\Anushaktinagar, Mumbai - 400094, India\\
% smishra@niser.ac.in \\ORCID: 0000-0002-9910-7291 }
% }

\maketitle

\begin{abstract}
Google app market captures the school of thought of users via ratings and text reviews. The critique’s viewpoint regarding an app is proportional to their satisfaction level. Consequently, this helps other users to gain insights before downloading or purchasing the apps. The potential information from the reviews can't be extracted manually, due to its exponential growth. Sentiment analysis, by machine learning algorithms employing NLP, is used to explicitly uncover and interpret the emotions. This study aims to perform the sentiment classification of the app reviews and identify the university students’ behaviour towards the app market. We applied machine learning algorithms using TF-IDF text representation scheme and the performance was evaluated on ensemble learning method. Our model was trained on Google reviews and tested on students’ reviews. SVM recorded the maximum accuracy(93.37\%), Fscore(0.88) on tri-gram + TF-IDF scheme. Bagging enhanced the performance of LR and NB with accuracy of 87.80\% and 85.5\% respectively.
\end{abstract}

Keywords:
Sentiment analysis, Machine learning, University students reviews, Google playstore apps.

\section{Introduction}
We are living in an era where technology and Internet have redefined social norms. There is no denying that mobile apps have changed every aspect of our lives completely\cite{M.Harman2012}. Irrespective of what we want or need to do; everything is simply at our fingertips, just by discovering the relevant apps and scrolling down their reviews and ratings posted by others. This helps in generation of profit for the developers, giving bug reports, request for new features, documentation of experience to analysts\cite{L.V.Galvis2013} and designers\cite{W.Maleej2011}. It gives information related to products, services, organizations, individual's issues, events, satisfaction or dissatisfaction with new features or business relevant information. Whether we are travelling\cite{Y.Blanco2010}, communicating\cite{P.Adinolfi2016}, watching movie\cite{T.T.Thet2010}, ordering products\cite{H.Cui2006}, performing bank transaction, there is app for everything, and so is the review.
New apps are rolling out every day with technical information available in the description; and ordered in terms of latest reviews, ratings, download strategy\cite{N.Seyff2010}. This helps in the qualitative and quantitative analysis of users' viewpoint for sizing and pricing strategy, technical claims, and features of apps. We need to efficiently analyze the technical, business and users' aspect of the app market because sometimes users' may intentionally or unintentionally leave a review that might be false regarding the technical claims made. Nevertheless, the observations from the entire app market may prove to be robust. However, the problem with the app market is that it stores a large number of reviews that takes longer computations and efforts. Secondly, the quality of reviews vary tremendously from helpful advice to a bad advice. Thirdly, filtering the negative and positive comments in the reviews are sometimes tricky.

Sentiment analysis help to mine the people’s opinions, sentiments, behaviors, emotions, appraisals and attitudes towards products or services, issues or events, topics\cite{Liu2015}. There are three types of people's opinions namely, positive, negative and neutral which identify the entire knowledge of the domain. %and hence identifies the overall subject of the domain. 
It is an integral part of the natural language processing(NLP) and helps in text mining and information retrieval.
In recent years, it has extended to fields like marketing, finance, political science, communications, health science. 
We can process the results and extract opinions from sentiment analysis and come to valuable conclusions\cite{L.V.Galvis2013}.

Machine Learning based techniques as well as lexicon based methods are used in sentiment analysis\cite{W.Medhat2014}. 
Lexicon based approach is an approach that considers the semantic order of the words and doesn't include labelled data. Dictionary is created manually and includes words and phrases in a document. Goal of sentiment analysis through machine learning approach deals with labelled data and helps to create model using supervised learning algorithms namely, Naïve Bayes(NB), support vector machine(SVM), and K‐nearest neighbor(KNN). \cite{A.Onan2016}

We had collected 10,841 Google app reviews with 13 fields to train our model\cite{data}. While for the sake of testing of our model, we collected 400 reviews with 6 fields from amongst the Utkal university students via local survey, department wise. This in turn, can be used as a measure for sentiment analysis and understanding local trends of the app market by other students. In addition, university student reviews can be utilized in the administrative related decisions.
Specifically, this paper presents the correlation between the university students reviews and the Google app reviews via an exploratory analysis and visualization of sentiment polarity, subjectivity versus other features like price, installs, type, size, category, ratings.

The contribution of our paper includes:
\begin{itemize}
    \item Use of multiple algorithms as well as text representation schemes for sentiment analysis on Google reviews dataset. 
    \item The text representation scheme namely, TF-IDF was implemented on uni-gram, bi-gram and tri-gram strategy.
    \item The supervised machine learning methods(such as, NB, SVM, logistic regression(LR), KNN, and Random Forest(RF)) was implemented on the text representation scheme and compared amongst each other with respect to its performance.
    \item The ensemble learning method(namely, bagging) was used with the classification algorithm namely, LR and NB and its performance was evaluated on text representation scheme. 
\end{itemize}

The organization of this paper comprises of five sections. Section 2 presents the related work in sentiment analysis. Section 3 describes the methods utilized in the paper. Section 4 introduces the experimental procedure, results and their analysis. Finally, Section 5 presents the concluding remarks of the study and future work.

\section{Related work}

Lima et al.\cite{Lima2015} have used majority voting scheme on the twitter dataset. They have combined the machine learning based paradigms and lexicon based methods. In their work, the tweets are a part of the labelled training data only when it consists of 5\% of words or emoticons otherwise it is considered a part of test data. Novak et al.\cite{Novak2015} have explained us about emoji based sentiment analysis and the 750 frequently used emojis were also analyzed in the twitter dataset. Lately, Onan et al.\cite{Onan2020} have collected instructors reviews from students for opinion mining using deep learning paradigm. The inferrence was GloVe with Recurrent Neural Network - Attention Mechanism algorithm has outperformed others.

While, Adekitan and Noma‐Osaghae\cite{Adekitan2019} have predicted about the performance of the university students using machine learning algorithms in their work. Linear and quadratic regression models were used for validation. Almasri et al.\cite{Almasri2019} have predicted the performance of students using ensemble tree‐based model. While Adinolfi et al.\cite{P.Adinolfi2016} evaluated student satisfaction on different learning e-platforms of online courses using sentiment analysis.

Recently, Jena\cite{Jena2019} have used machine learning algorithms(namely, NB, SVM, entropy classifiers) along with conventional text representation schemes(namely, uni-gram, bi-gram, tri-gram) on students data for obtaining the sentiment polarity.

The past reports on NLP in sentiment analysis does not capture any of the comparison between the machine learning algorithms. So, our literature survey was concentrated on instructors review paper approach\cite{Onan2020} that threw light on multiple combinations of machine and deep learning algorithms. The latest trend observed from our paper is that it not only emphasises on count vectorizer method of splitting the data set but we also aggregated a  new university data set. Focus is on data analysis along with modeling.  

\section{METHODOLOGY}

This section briefly describes the methods used in our study. Fig.1 explains the proposed methodology.

\subsection{Data Sources}

The corpus, Google reviews, was collected in .csv format\cite{data}. There were 9659 apps, 33 categories, 115 genres in the dataset. The columns of the dataset are as follows app(name), category(app), rating(app), reviews(user), size(app), installs(app), type(free/paid), price(app), content rating(everyone/teenager/adult), genres(detailed category), last updated(app), current version(app), android version(support).

In this study, the aim is to understand the trend of Google app market and comparing it with test data i.e., analyzing the students' behavior towards the Google app market. 
So, we had collected the real-life data from the Utkal university students, department wise. 
The survey was entirely voluntary in nature and no incentives were offered to perform the survey. If university students did not wish to participate, then they were excluded from the survey.
The reviews regarding the frequently used apps, were gathered via a survey for a month. The survey was made on an online platform via a Google form. One student from one department could at max list out seven frequently apps used on their device. The fields included department, app(name), reviews, ratings, type. Students rated the apps on a 5-point scale where ratings below 3 were labeled as "negative" and that with 3 or greater than 3 is considered "positive". 

Data cleaning, text pre-processing techniques were carried out on the dataset to build efficient learning models and enhance the overall performance. These included, removing  missing data, dropping NA values, removing punctuation, tags, special characters URLs, emojis, digits, filtering stop words. Tokenization, noise removal, spelling correction, stemming, lemmatization\cite{GAMiller1995} were highlighted in the empirical analysis.
After cleaning, there were 9,360 and 380 records in the training and test data set respectively.

Table-I presents some sampled test data reviews along with the sentiment characteristics. TextBlob package in Python helps in calculations for sentiment analysis. A sentiment score determines how negative or positive the entire text analyzed is. 
For eg., the phrase “not a very great app” has a polarity of about -0.3, implies it is slightly negative, and a subjectivity of about 0.6, implies it is fairly subjective.

\begin{table}[h!]
\centering
\caption{Sample students' reviews and Sentiment Characteristics from Students' dataset, Orientation - determines positivity, negativity or neutality of sentence, Polarity - helps identify the sentiment orientation, Subjectivity - defines person' opinions, emotions or judgment; ranging from 0.0 (objective) to 1.0 (subjective) }
% \begin{tabular}{  p{5em} | p{2cm} | p{2cm} | p{1cm} | p{2cm} }
\begin{tabular}{  p{10em} | p{4em} | p{4em} | p{3em} | p{3em} }
\hline
Students' review & App & Orientation & Polarity & Subjecti- vity \\
\hline
It's helpful to learn at home.Highly recommendable & Unacademy & Positive & 0.04 & 0.135 \\
It's amazing and works well.  & PhonePay & Positive   & 0.3 & 0.725 \\
Horrible. Keeps crashing my phone. & Subway Surfers & Negative & -0.104 & 0.43 \\
It' annoying due to adds.   & JioSaavn & Negative & -0.033 & 0.388 \\
Very well designed. Many updates present. & WPS Office & Positive &1 &0.75\\
\hline
\end{tabular}

\label{TABLE-I}

\end{table}

\subsection{Text Representation Schemes}

Bag-of-words paradigm\cite{G.Hackeling2017} is a very commonly used technique to represent all the unique words occurring in the documents. The occurrences of the terms in a document is noted while the order and the sequence of words is not considered. This scheme helps in feature extraction from text documents. The three weighted schemes frequently utilized are based on bag-of-words model i.e., Term Presence(TP), Term Frequency(TF) and TF-IDF(Term Frequency-Inverse Document Frequency). 
 
TF-IDF scheme is an improvement over TP\cite{J.Brownlee} and uses a normalizing aspect for computations. TF-IDF basically combines two metrics, namely TF and IDF. It helps in ranking the queries in search engines and used widely in information retrieval and text mining. It is used to weight words according to their importance.

Mathematically, TF-IDF is defined as: 
\begin{equation}
TF-IDF = TF(w,D)* log(C/df(w))
\end{equation}
Here,  $TF-IDF(w, D)$ maps the TF-IDF score for a word $w$ in document $D$.
Therefore, it will score higher if the term is not common.
TF(w,D) is the frequency of term in a given document(synonymous with bag of words).
IDF measures the significance of the word in the corpus of documents. Given a corpus $C$, the number of documents divided by the frequency of word in the document $w$. followed by the log transform gives IDF.  
Highly occurring words across many documents will have a lower weight, and otherwise would have a higher weight.
% Words that are frequently used in many documents will have a lower weighting while infrequent ones will have a higher weighting.

N-gram model is a collection of words from a text document in which the the words are contiguous and occur sequentially. They may be in the form of phrases or group of words. In n-gram model:
\begin{itemize}
  
\item when n is one (order is one) i.e., it consist of one word, therefore it is termed as an uni-gram model.
\item Similarly, bi-gram model indicates n is two (order is two) i.e, it consists of two words. 
\item Tri-grams indicates n is three (order is three) i.e., it consists of three words and so on. The n-gram model is a supplement of the bag-of-words model.
\end{itemize}
In this study, we performed an experiment on the Google apps corpus based on three n-gram model and TF-IDF, and obtained three different configurations.

\subsection{Machine Learning Algorithms}
If a machine learning algorithm is trained on a dataset tagged with labels, it is called supervised learning. Labelling basically marks the output on the input parameters. % Supervised learning is used when the model gets trained on a labelled dataset. Labelled dataset is one which has both input and output parameters. 
We trained our model on different N-gram models(i.e., uni-gram, bi-gram, and tri-gram models) and TF-IDF based weighting scheme using google apps reviews. As a result, three different feature sets were obtained. In the next subsections, we briefly describe the details of the five most frequently used classifiers in sentiment analysis.

LR\cite{T.Hastie2009} helps to solve a classification problem by analyzing a dataset where the outcome depends on one or more independent features. It is a linear algorithm and the underlying technique is quite similar to Linear Regression. The term “Logistic” is taken from the Logit function that is used in classification. The idea is to come up the model that best describes the relationship between the outcome and a set of independent variables. The dependent variable is binary , i.e., it only contains data coded as 1 (TRUE, success) or 0 (FALSE, failure).

SVM\cite{V.Vapnik1998} is a supervised machine learning algorithm. It helps to solve classification and regression problems. In SVM, the data points are plotted in N-dimensional space where N denotes number of features and a hyper-plane is found to differentiate the data points. However, this algorithm is computationally expensive but is used when the number of dimensions is high with respect to the number of data points.

NB\cite{D.Lewis1998} is a probabilistic classifier which uses Bayes theorem. The object with similar features are grouped in one class while others in a different class based on certain probability. In this method, there is a strong independence assumptions between the features. It requires a small training data for classification, and all terms can be pre-computed thus, classifying becomes easy, quick and efficient.

KNN\cite{D.W.Aha1991} solves the classification problem by assigning the object to a class by a plurality vote from its k(positive integer) neighbors. While in regression, the output value for an object is the aggregate values of k nearest neighbors. KNN captures the idea of similarity amongst the object with respect to its neighbors in terms of distance, proximity, or closeness.

RF\cite{L.Brieman2001} widely uses bagging, random subspace methods, ensemble learning paradigms. For the purpose of classification decision trees are used. It decomposes the training dataset set based on certain conditions on attribute values using random sampling. The data is divided recursively until the leaf node is left with minimum amount of records using random subset features.

\subsection{Ensemble Learning Methods}

The base estimators are built on a given learning algorithm and their predictions can be combined to improve the robustness and performance over single estimator\cite{A.Onan2017}. It includes averaging and boosting methods. 
In averaging methods, several estimators are built independently and their predictions are averaged. The combined estimator performs better due to the reduced variance. Examples Bagging method, Forests of randomized trees.
By contrast, boosting methods\cite{Y.Freund1996}, reduces the bias of the combined estimator by building the base estimators sequentially. Several weak models can be combined to generate a powerful ensemble. Examples AdaBoost, Gradient Tree Boosting.
In this study, we have considered Bagging method and the rest of the section describes it.

Bagging is also called as Bootstrapped Aggregation\cite{L.Breiman2001}. It is used for predictive modeling (CART). 
Random subsets of data are drawn from the training dataset with replacement, and a final model is produced by averaging result from several models. 
One popular way of building Bagging Models is by combining several DecisionTrees with reduced bias that increases the model’s prediction than individual Decision Trees. Averaging ensembles with bagging techniques like RandomForestClassifier and ExtraTreesClassifier reduces the variance, avoids over-fitting and increases the model's robustness with respect to small changes in the data.

\begin{figure}
    \centering
    \includegraphics[width=0.4\textwidth]{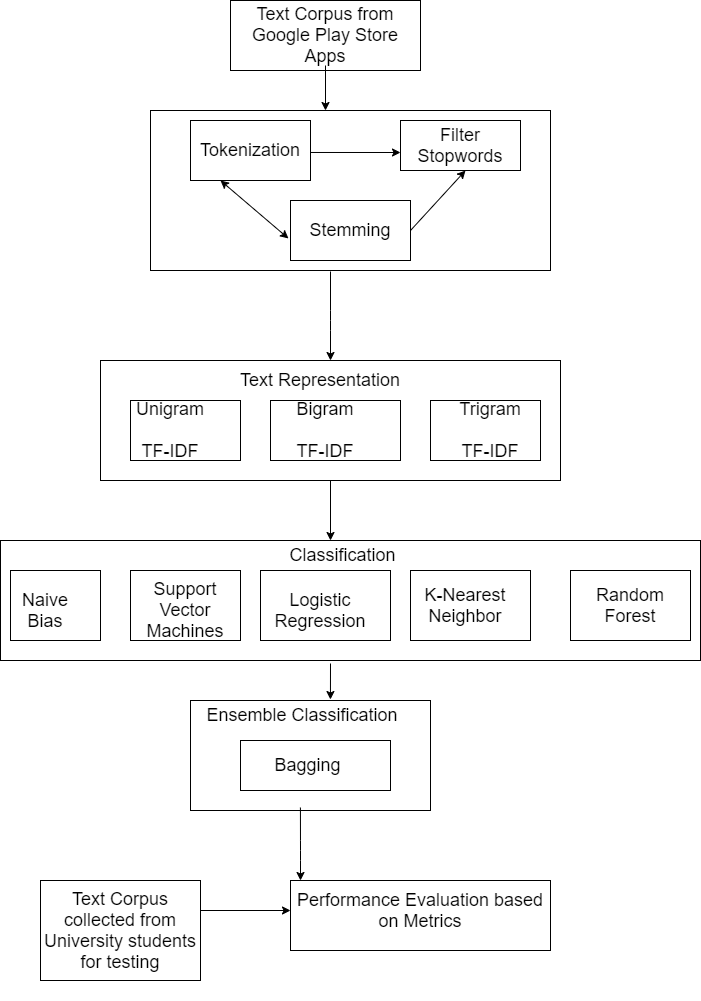}
    \caption{Proposed Methodology}
    \label{fig:my_label}
\end{figure}

\subsection{Result Evaluation Metrics}

This section briefly discusses about the metrics used in this study for the result computation.

Precision: It measures correctness of a classifier. It is the proportion of number of precisely extracted opinions to the total number of extracted opinions.
\begin{equation}
precision=\frac{TP}{TP+FP}
\end{equation}

Recall: It measures the sensitivity of a classifier. It is defined as the proportion of number of precisely extracted opinions to the total number of annotated opinions.
\begin{equation}
recall=\frac{TP}{TP+FN}
\end{equation}

F-Measure is other commonly used measure. Precision and recall are combined to give a single criterion called as F-measure. The harmonic mean of precision and recall computes F measure. It rates a system with one unique rating.
\begin{equation}
F-measure=\frac{2*precision*recall}{precision+recall}
\end{equation}

Accuracy is commonly used performance measure in supervised learning techniques. It is the ratio of truly predicted observation to the total number of observations.
\begin{equation}
accuracy=\frac{TP+TN}{TP+TN+FP+FN}
\end{equation}

Here, TP refers to true positive, i.e., the positive tuples that were faultlessly characterized by the classifier.
TN refers to true negative i.e., the negative tuples that were faultlessly characterized by the classifier. 
FP refers to false positive i.e., the negative tuples that were inaccurately characterized by the classifier as positive.
FN refers to false negative i.e., the positive tuples that were untruly characterized by the classifier as negative.

\section{Experiments and Results Analysis}

The empirical analysis was done to train our model on the Google reviews dataset and test our model on Utkal university students reviews dataset. The platform used was Python and the contribution of paper is two-fold.
Firstly, an exploratory analysis on our training dataset i.e., the Google app reviews and compared the results with the university students dataset through visualization was done. This helps us to get a glimpse of the behaviour of students towards the distribution of the app market.
Research Questions(RQ) were investigated to understand the correlation between the price, popularity and ratings of apps by the students when compared with that of training dataset. These are presented in Table-II. Secondly, we performed an experiment to train our model using the classification algorithms on the conventional text representation schemes. We used evaluation metrics to generate useful intuitions from our corpus. The three different configurations so obtained are listed in Table-II and Table-III and the conclusions are as follows:

\begin{table*}[ht]
\centering
\caption{Assessing the Research Questions.}
\begin{tabular}[t]{p{3em} p{15em} p{5em}  p{20em}}
\hline
Serial & RQ & Figure & Answer \\
\hline
RQ1
& Do the apps which get a higher rating in the training dataset tend to be more popular among the students as well? 
& 
Fig. \ref{rq1} and and Fig\ref{xx}
& 
The Google app market breakdown showed prominent downloads in Social and Games categories. On the contrary, Weather and Comics were of least interest among students.
The average ratings shooted up to 4.17 across major categories.
Interestingly, Shopping, Food and Drinks, News and Magazine are also catching up.
Expensive apps may make students disappointed, if they are not good enough and consequently get low ratings.
Students from Mathematics and Sanskrit department participated fairly well while Women Studies and Geography showed least participation. Other departments showed an acceptable participation.\\

RQ2
&
 Do the priced and free apps get the same ratings and popularity from the students as compared to the training dataset?
 &
 Fig \ref{rq2}
 &
This jointplot visualization depicts the sizing strategy(small vs huge). We got a clear conclusion that, small sized app(0-60 Mb) are predominant for downloads among students. This enhances the ratings. Average rating turning out to be 4-5. On the contrary, larger apps have least ratings and less preferable.\\

RQ3
&
 What is the correlation between price, rating, popularity amongst the university students when compared with the training dataset?
 
&Fig \ref{rq3}
&The installs and reviews are positively correlated amongst students. While, installs and pricing are negatively correlated.\\

RQ4
&
What is the sentiment polarity could be analyzed by the reviews of students when compared with the training dataset?
&
Fig \ref{rq4}
&
The scatter plots are heavily clustered towards positive side rather than on negative. Specifically, we can say students weren't so harsh while give reviews, instead gave genuine and lenient feedback. \\

RQ5
&
How size of apps affect the installs amongst the students as compared to the training dataset?
&
Fig \ref{rq5}
&
The points in the jointplot are heavily clustered where the price for apps are 0. This gives us an inference that students prefer free apps rather than paid and an average rating between 3.5 to 5 is shown.\\

RQ6
&
What confusion matrix can be obtained from students' reviews as compared to the training dataset?
 &
Fig \ref{rq7}
&

The confusion matrix when applied with LR gave us an accuracy of 90.8\%.\\  
\hline
\end{tabular}
\label{TABLE:II}
\end{table*}

\begin{figure*}
    \centering
    \includegraphics[width=0.8\textwidth]{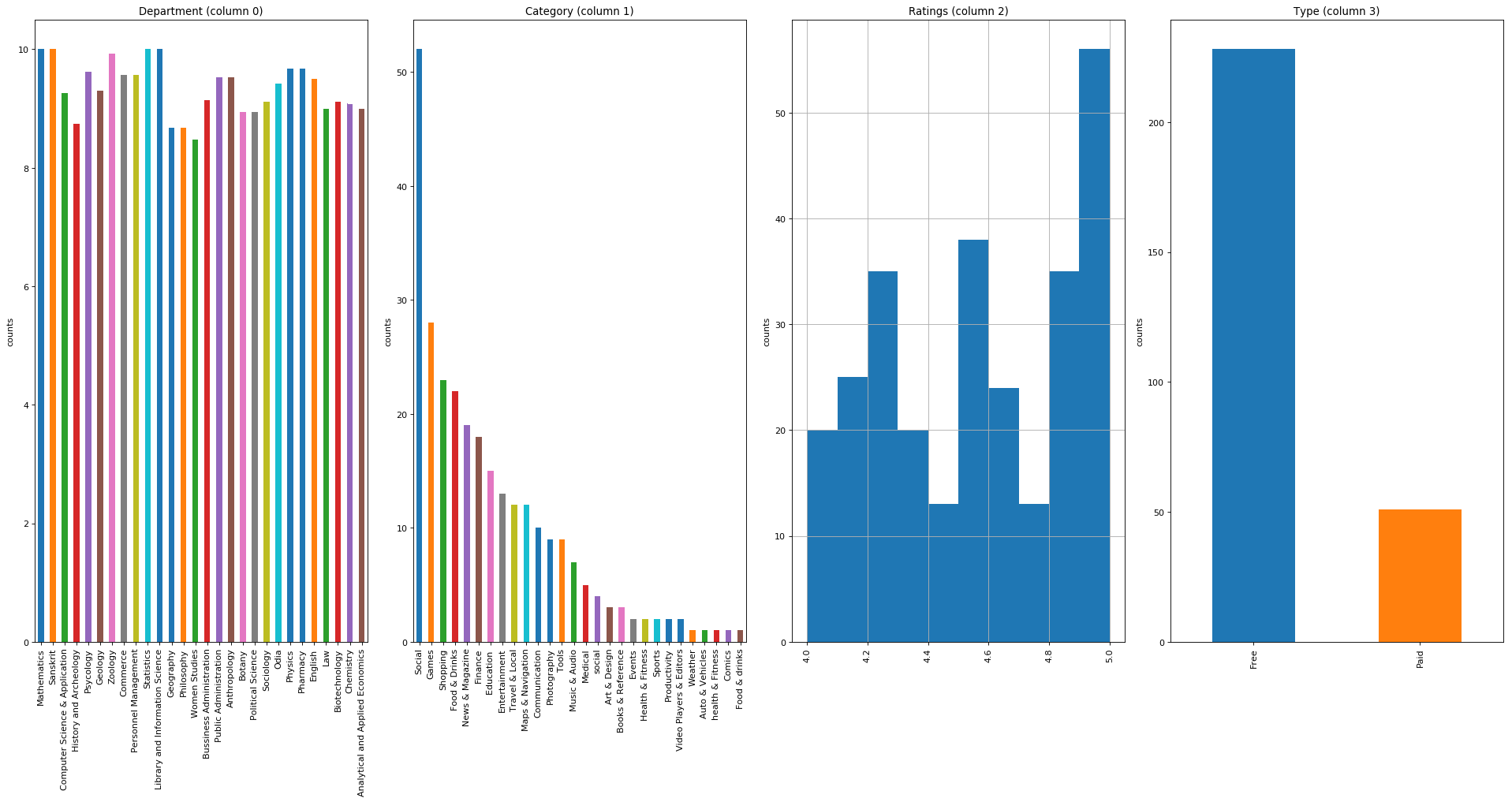}
    \caption{Distribution of counts over Category, Ratings, Price}
    \label{rq1}
\end{figure*}

% \begin{figure}
%     \centering
%     \includegraphics[width=0.5\textwidth]{RQ11.jpg}
%     \caption{Distribution of counts of over Department}
%     \label{fig:rq1}
% \end{figure}

\begin{figure*}
     \centering
     \begin{subfigure}[b]{0.3\textwidth}
         \centering
         \includegraphics[width=1\textwidth]{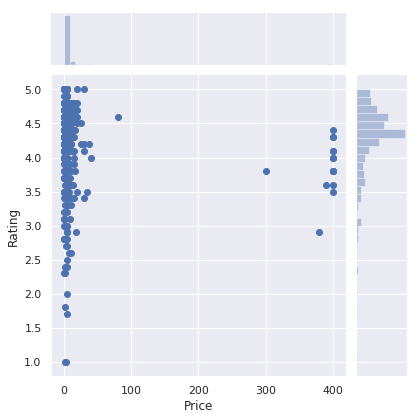}
         \caption{Distribution of Price over Ratings}
         \label{rq2}
     \end{subfigure}
     \hfill
     \begin{subfigure}[b]{0.3\textwidth}
         \centering
         \includegraphics[width=1\textwidth]{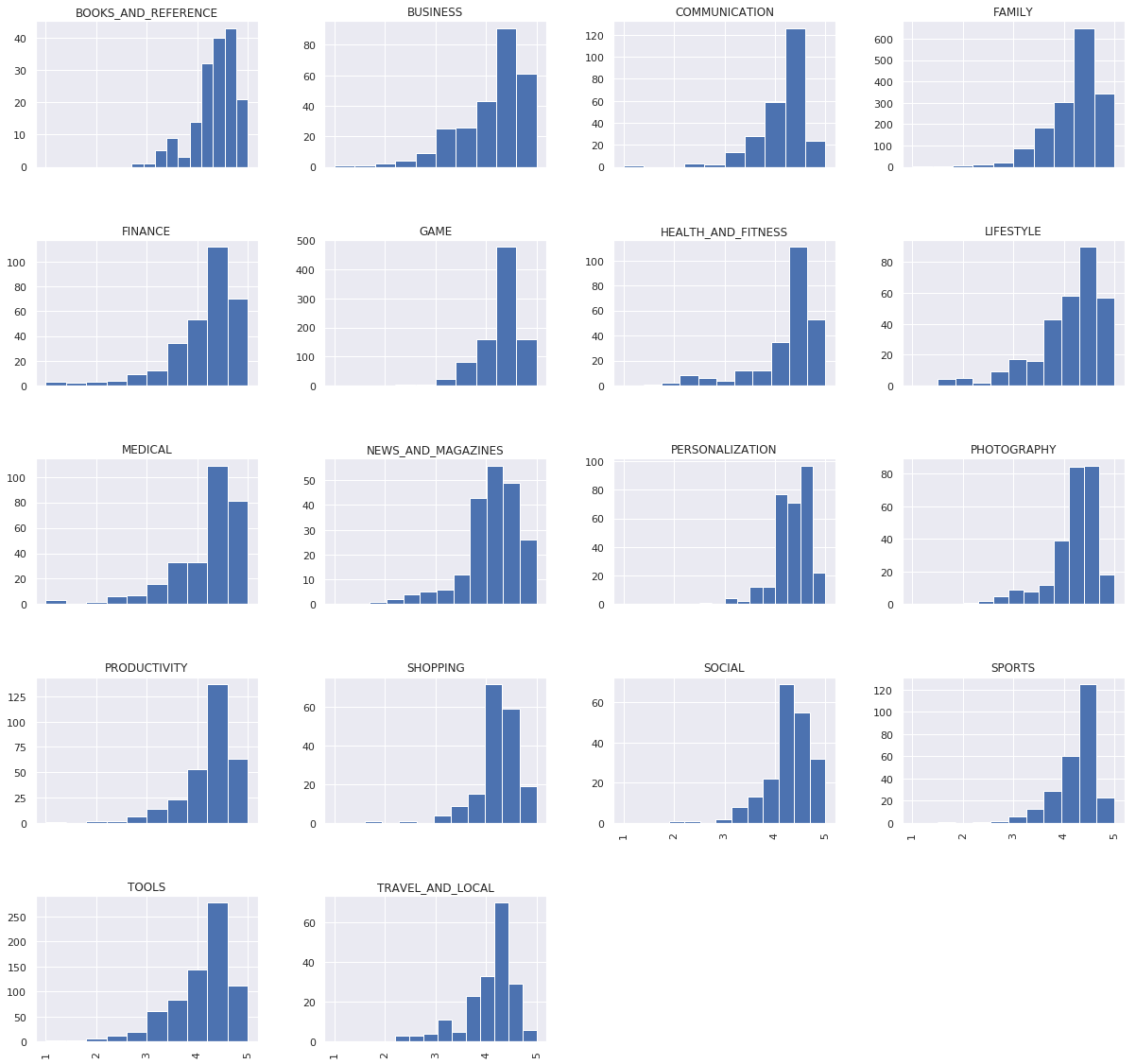}
         \caption{Distribution of Ratings over Category}
         \label{xx}
     \end{subfigure}
     \hfill
     \begin{subfigure}[b]{0.3\textwidth}
         \centering
         \includegraphics[width=1\textwidth]{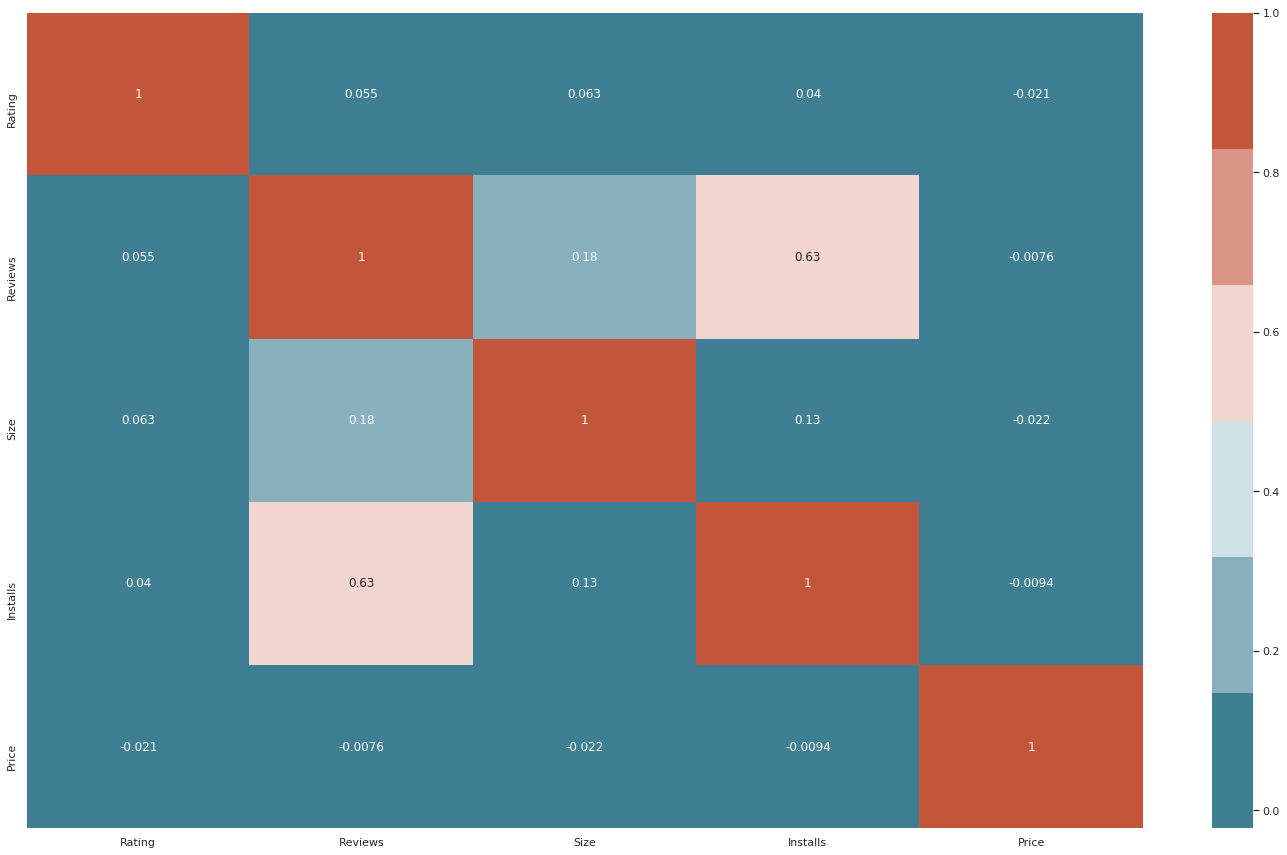}
         \caption{Correlation between training and test data set}
         \label{rq3}
     \end{subfigure}
     \hfill
     \begin{subfigure}[b]{0.3\textwidth}
         \centering
         \includegraphics[width=1\textwidth]{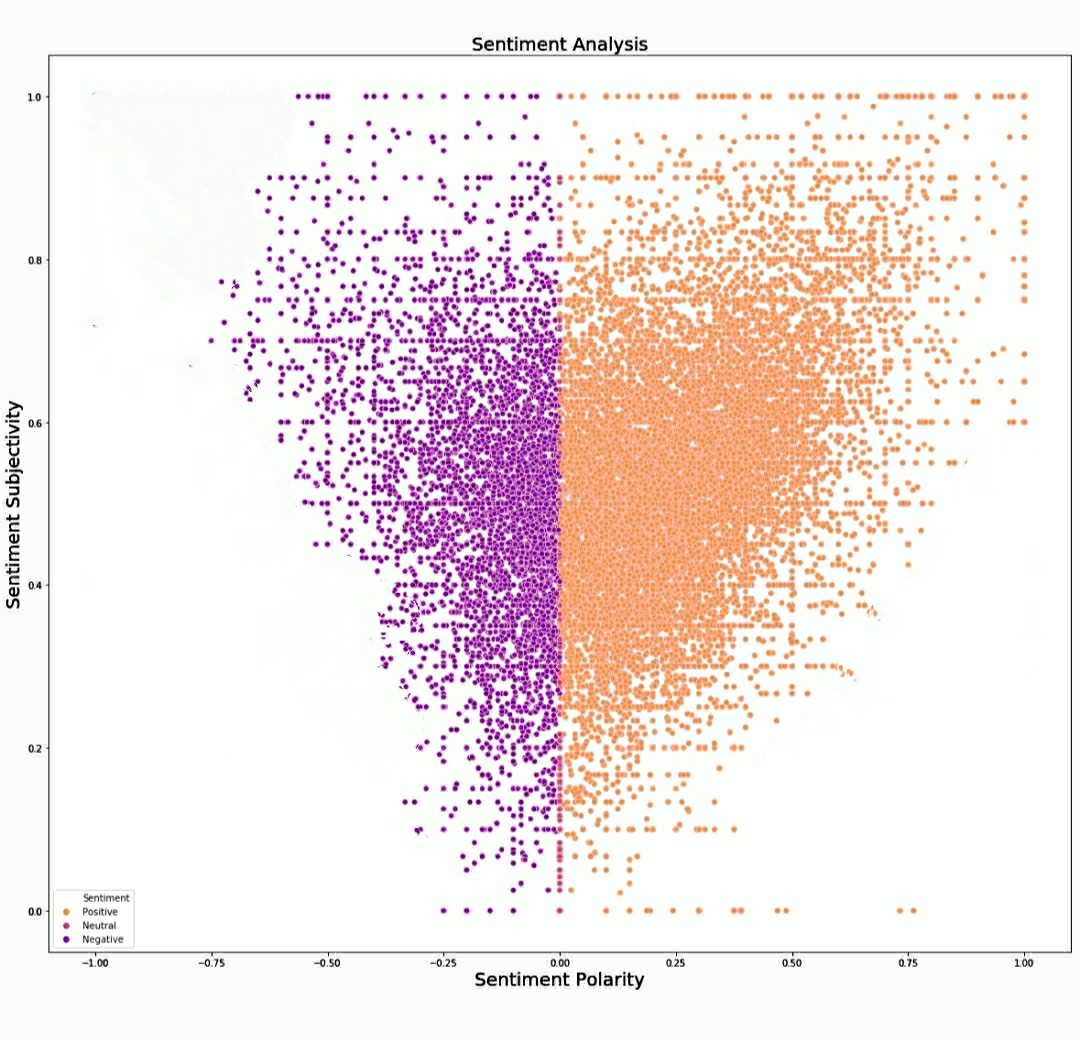}
         \caption{Distribution of Sentiment Subjectivity over Sentiment Polarity}
         \label{rq4}
     \end{subfigure}
     \hfill
     \begin{subfigure}[b]{0.3\textwidth}
         \centering
         \includegraphics[width=1\textwidth]{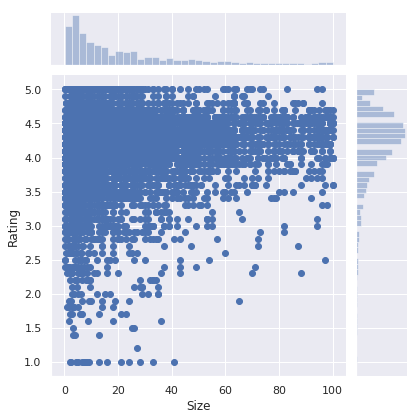}
         \caption{Distribution of Size over Installs}
         \label{rq5}
     \end{subfigure}
     \begin{subfigure}[b]{0.3\textwidth}
         \centering
         \includegraphics[width=1\textwidth]{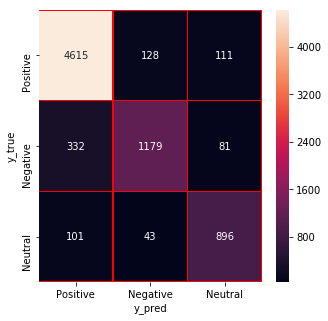}
         \caption{Confusion Matrix}
         \label{rq7}
     \end{subfigure}
     
        \caption{Visualization of Research Questions}
        \label{qr}
\end{figure*}

\begin{table}[ht]
\centering
\caption{Accuracy values obtained by Machine Learning algorithms}
\begin{tabular}[t]{lccc}
\hline
&Uni-gram+TF-IDF &Bi-gram+TF-IDF &Tri-gram+TF-IDF\\
\hline
SVM & 92.89 & 93.41 & 93.37 \\
KNN & 91.01 & 90.90 & 88.39\\
LR & 84.08 & 84.61  & 84.48\\
RF & 83.42 & 85.11 & 84.16\\
NB & 80.00 & 82.14 & 82.21\\
LR(Bagging) & 86.50  & 86.50 & 87.88\\
NB(Bagging) & 85.50 & 85.11 & 84.00\\

\hline
\end{tabular}
\label{TABLE:III}
\end{table}%

\begin{table}[ht]
\centering
\caption{F-values obtained by the Machine Learning algorithms}
\begin{tabular}[t]{lccc}
\hline
&Uni-gram+TF-IDF &Bi-gram+TF-IDF &Tri-gram+TF-IDF\\
\hline

SVM & 0.89 & 0.89 & 0.88\\
LR & 0.86 & 0.85 & 0.85\\
KNN & 0.69 & 0.68 & 0.70\\
RF & 0.68 & 0.61 & 0.62\\
NB & 0.72 & 0.62 & 0.63\\
LR(Bagging) & 0.87  & 0.86 & 0.86\\
NB(Bagging) & 0.75 & 0.76 & 0.76\\

\hline
\end{tabular}
\label{TABLE:IV}
\end{table}%

\begin{itemize}
\item Algorithms like SVM and KNN performed best on this dataset on all three representations uni-gram, bi-gram, tri-gram with TF-IDF featurization. SVM performed the best in terms of accuracy(93.41\%) and F-score(0.89).
\item While KNN achieved an accuracy(91.01\%) and F-score(0.88) on uni-gram+TF-IDF featurization.
\item LR gave an accuracy of 84.08\%  and F-score of 0.86 on unigram+TF-IDF scheme. It is an average algorithm as compared to others and didn't take a lot of time to train. It performed fairly well on our dataset when applied with bagging. The accuracy of 87.8\% and F-score of 0.87 so obtained outperformed NB and RF.
\item NB proved to be very slow with least accuracy(80\%) and F1-score(0.72) amongst others algorithms and was not that good for this dataset. However, when applied with Bagging an accuracy(85.50\%) and F-score(0.72) was achieved and outperformed LR and RF. 
\item RF performed fairly well on our dataset with accuracy(85.11\%) and F-score(0.68) on uni-gram+TF-IDF scheme. It turned out to be an average algorithm in our study.
\end{itemize}
Code for the experimental analysis will be available at https://github.com/smlab-niser/Google-Reviews-Sentiment-Analysis

\section{Conclusions and Future work}

The objective of this paper was to efficiently model the sentiment of the users using the Google reviews dataset and find the university students' behavior towards the Google app market. 
Usually, k-fold cross validation technique is used for testing. Not much research has been done using students' reviews for testing. So, we had collected the real-life dataset from Utkal university students proposed model. Machine learning paradigm was efficiently employed to perform the sentiment analysis. Five commonly used classification algorithms namely NB, SVM, KNN, LR, RF were used for performance comparision. Bagging, an ensemble method, was employed to intensify the predictive performance of the classifier. In this study, amongst the classification algorithm, SVM outperformed others in terms of accuracy(93.41\%) on the TF-IDF+bi-gram feature, while NB underperformed with an accuracy(80\%). In terms of F-score, SVM and LR performed significantly well.
% Despite having favourable results, there are certain limitations of this paper namely the survey was localized only to a particular university and we did not get a 100\% participation of university students.

Future scope could be, empirical analysis on TP and TF-based representation in conjugation with uni-gram, bi-gram and tri-gram model respectively. Ensemble methods like random subspace and boosting can be employed to study the predictive performance. Analysis of word embeddings(word2vec and GloVe) could also be explored. Eventually, we could expand our dataset by extending our online survey in other universities and more students within the range of the city. Identification of influential reviewers and text summarization could be one of the grounds for future research.

\section*{Acknowledgement}

We would like to dedicate this work of ours to one of our favourite Professor and former HoD Sir (late) Dr. B.K.Ratha, Department of Computer Science and Applications, Utkal University. A life full of endless possibilities cut short in a moment at an age of 54, due to deteriorating health. 

He encouraged us all the way and whose encouragement has made sure that "we give it all it takes to finish that which we have started". He is missed everyday because he has left a void never to be filled in our lives. His care and concern for us knew no bounds. Our respect for him can never be quantified. May he find peace and happiness in Paradise!

\end{document}